\documentstyle[11pt,aaspp4,tighten,flushrt]{article}

\lefthead{Sakamoto et al.}
\righthead{Counterrotating Disks in Arp 220}

\newcommand{\etal}{\mbox{\it et al.}}
\newcommand{\Msol}{\mbox{$M_{\odot}$}}
\newcommand{\Lsol}{\mbox{$L_{\odot}$}}
\newcommand{\kms}{\mbox{km s$^{-1}$}}
\newcommand{\beam}{\mbox{beam$^{-1}$}}

\newcommand{\Halpha}{\mbox{H$\alpha$} }
\newcommand{\tm}[1]{\tablenotemark{#1}}

\begin{document}

\title{Counterrotating Nuclear Disks in Arp 220}

\author{
K. Sakamoto\altaffilmark{1}, 
N. Z. Scoville\altaffilmark{1},
M. S. Yun\altaffilmark{2},
M. Crosas\altaffilmark{3},
R. Genzel\altaffilmark{4},
and
L. J. Tacconi\altaffilmark{4}
}

\altaffiltext{1}{California Institute of Technology, 105-24, Pasadena, CA 91125;
e-mail(KS): ks@astro.caltech.edu}
\altaffiltext{2}{National Radio Astronomy Observatory,
	P.O. Box 0, Soccoro, NM~~87801-0387}
\altaffiltext{3}{Harvard-Smithsonian Center for Astrophysics, 60 Garden Street, 
Cambridge, MA 02138}
\altaffiltext{4}{Max-Planck-Institut f\"{u}r Extraterrestrische Physik, D-85740, Garching, Germany}

\begin{abstract}
	The ultraluminous infrared galaxy Arp 220 has been observed 
at 0\farcs5 resolution in CO(2--1) and 1 mm continuum using 
the newly expanded Owens Valley Millimeter Array. 
The CO and continuum peaks at the double nuclei 
and the surrounding molecular gas disk are clearly resolved.
We find steep velocity gradients across each nucleus
($\Delta V \sim 500$ \kms\ within $r= 0\farcs3$) 
whose directions are not aligned with each other and with
that of the outer gas disk.
We conclude that the double nuclei have
their own gas disks ($r \sim 100$ pc).
They are counterrotating with respect to each other 
and embedded in the outer gas disk 
($r \sim 1$ kpc) rotating around the dynamical center of the system.
The masses of each nucleus are
$M_{\rm dyn} \gtrsim 2\times 10^{9} \Msol$ based on the CO kinematics.
Although there is no evidence of an old stellar population in the optical 
or near infrared spectroscopy of the nuclei (probably due to the much
brighter young population), it seems likely that these nuclei were
'seeded' from the pre-merger nuclei in view of their counterrotating
gas kinematics. The gas disks probably constitute a 
significant fraction ($\sim$50\%) of the mass in each
nucleus. 
The CO and continuum brightness temperatures imply that 
the nuclear gas disks have high area filling factors ($\sim$ 0.5--1)
and have extremely high visual extinctions ($A_{V} \simeq 1000$ mag).
The molecular gas must be hot ($\geq 40$ K) 
and dense ($\geq 10^{4-5}$ cm$^{-3}$),
given the large mass and small scale-height of the nuclear disks.
The continuum data suggest that the large luminosity 
(be it starburst or AGN) must originate  
within $\lesssim$ 100 pc of the two nuclear gas disks
which were presumably formed through concentration of gas from the progenitor outer galaxy disks.
\end{abstract}


\vspace{1 cm}
\begin{center}
Accepted for publication in {\it The Astrophysical Journal (part 1)}
\end{center}

\section{Introduction}
	Arp 220 is the prototype ultraluminous infrared galaxy with 
$L_{\rm 8-1000 \micron} = 1.4\times 10^{12} \Lsol$ (Soifer \etal\ 1987). 
The galaxy is thought to be in the final stage of merging since optical images
show extended tidal tails (Arp 1966) and radio and NIR
imaging reveal double nuclei with projected separation $\sim 300$ pc (Norris 1985; Graham \etal\ 1990).
Arp 220 also has a massive concentration of molecular gas ($\sim$10$^{10} \Msol$) 
within the central kiloparsec (Scoville \etal\ 1986).
These characteristics are similar to those of 
many of the ultraluminous infrared galaxies which often show morphological
evidence of strong interactions or merging (Sanders \etal\ 1988) and have large 
molecular gas masses ($\sim$10$^{10} \Msol$; Sanders, Scoville, \& Soifer 1991; 
Solomon \etal\ 1997). 
The concentration of molecular gas in the nucleus is in accord with 
gas dynamical simulations of galaxy mergers 
(e.g., Barnes \& Hernquist 1991).

	High-resolution observations of molecular gas and dust continuum 
can provide crucial information on gas dynamics and thermal structure
in the nuclear region of Arp 220.
Arcsecond resolution observations
by Scoville, Yun, \& Bryant (1997; hereafter SYB) 
revealed a kiloparsec-size molecular gas disk rotating around 
the double nucleus.
Peaks in the CO line and mm-wave continuum at the two nuclei in the kpc-scale disk
were suggested by Scoville, Yun, \& Bryant and confirmed by Downes \& Solomon (1998; hereafter DS).
High resolution near infrared imaging with NICMOS on HST shows an 
area of extremely high dust obscuration south of the western nucleus
which is interpreted as a thin opaque, dust disk embedded in the 
central star cluster (Scoville \etal\ 1998). 
The formation of these nuclear disks during a galactic merger 
may occur naturally as a result of the high dissipation rates in the
dense gas, but their role in the promotion of nuclear starbursts
and feeding an pre-existing AGN (specifically the source of the large FIR luminosity) remains very ill-defined. However, given the fact that the molecular gas in the 
nucleus is by far the largest reservoir 
of ISM available for both star formation and 
nuclear accretion, it is likely that the nuclear disks play a central role in the 
nuclear activity. In this paper, we report 0\farcs5 resolution 
observations of molecular gas and dust in Arp 220 with the goal 
of defining better both the disk structure and the kinematics
of the nuclei. 
We adopt the distance of 77 Mpc for Arp 220, at which 1\arcsec\
corresponds to 373 pc.

\section{Observations}
	Aperture synthesis observations were made on 19 March 1998 
using the Owens Valley millimeter array. 
The six 10 m antennas were deployed in the new U (ultra-high) 
configuration that has 400 m (E--W) and 330 m (N--S) baselines.
Two SIS receivers were used simultaneously to observe the CO(J=2--1)
line and the 1.3 and 3.0 mm continuum. Gains and passbands were calibrated 
with observations of the quasars 1611+343 and 3C273, respectively. The  
flux scale for the quasars was established by observing Neptune\footnote{
Flux observations were made in a separate track to avoid the known CO absorption
in Neptune atmosphere. 
} and is accurate to $\sim 10$ \%.
The derived flux densities of 1611+343 were 1.4 Jy at 1.3 mm and 2.3 Jy at 3 mm. 
The digital spectrometers were configured identically to SYB 
in order to span the very broad CO line ($\Delta V = 900$ \kms)
and to facilitate combination of this data with the earlier 1\arcsec
resolution data.
Continuum data were recorded from upper and lower sidebands 
for both 1 and 3 mm in a 1 GHz bandwidth analog continuum correlator.

	The raw data were calibrated with the OVRO/MMA 
software (Scoville \etal\ 1993) and maps were processed using the NRAO AIPS
package. 
The continuum in the image sideband was subtracted from the line data 
in the $u$-$v$ domain. 
Twenty-one channel maps, each of 30.7 \kms\ width, were CLEANed and then 
combined for the moment maps. 
We have made two sets of maps; one from only the U-array data
and the other from the U-array data plus the L and H-array data 
taken by SYB (which we call LHU-array maps).
The former maps filter out most of extended emission while 
the latter retains it; the minimum physical baseline lengths 
are 100 m and 15 m, respectively.

\section{Results}

CO channel maps from the U-array data are shown in Figures~\ref{channel}
and \ref{chanLHU} respectively.
Figure~\ref{total} shows the integrated intensity 
and mean velocity maps from U- and LHU-array data.
The main features in the maps are largely consistent with those
in SYB and DS, including the CO and continuum peaks at
the east and west nuclei, which we call E and W, and the
CO disk about $5''$ in extent with a velocity gradient across it, 
approximately along the major axis.

Measured parameters of the CO and continuum emission are summarized in Table~1.
The separation and position angle of the two nuclei, 
0\farcs90 and 101\arcdeg, are in excellent agreement with those at 
other wavelengths (Norris 1985; Graham \etal\ 1990; Scoville \etal\ 1998).
Based on the single-dish line and continuum fluxes reported by 
Radford, Solomon, \& Downes (1991)
and Carico \etal\ (1992) respectively, we conclude that the LHU maps recover 
the entire CO and continuum flux from Arp 220. The
U-array maps also fully recover the entire continuum flux (within the
$\sim 15$ \% calibration 
uncertainties) but only 30 \% of total CO line flux.

\section{Continuum Emission}
	The 1 mm continuum is predominantly thermal emission 
from dust (cf. Scoville \etal\ 1991), i.e. the long wavelength 
tail of the far infrared emission component containing the bulk of the FIR luminosity of Arp 220. 
The continuum map in Figure~\ref{total} clearly shows 
that most of the dust emission arises from the two nuclear components 
with less than 20 \% contained in a more extended source 
(as noted above this map recovers the entire 1.3 mm continuum flux). 
The continuum flux ratio between the East and West nuclei is 
about $1:2$ at both 1 and 3 mm.
The spectral slope of each nucleus between 1.3 and 3.0 mm
is  $S_{\nu} \propto \nu^{3.3\pm 0.3}$
after subtraction of synchrotron emission 
as discussed in Scoville \etal\ (1991).
This long-wavelength spectrum is consistent with the emission from optically thin dust with an emissivity $\epsilon \propto \nu^{1.3}$ at $T \simeq 42$ K
(cf. Scoville \etal\ 1991 who modeled the total (E+W) flux).

	We find no evidence for the extended continuum emission reported by DS. 
Their 1.3 mm fluxes attributed to the E and W nuclei are only 45 \% and 60 \%
of our values but their total flux is 84 \% of that reported by us.
The smaller fluxes from the nuclei in DS are very likely due to their subtraction of an 
extended `background' or disk component but the method of 
separation is not described by them. Their published map shows no obvious 
extended component and the total flux measured by us in the E and W nuclei 
is 208 mJy.  Carico \etal\ (1992) report a single dish flux of 226$\pm10$ mJy at 1.25 mm which translates to 213 mJy assuming an emissivity spectral index of 1.3. 

	The continuum sources are extremely compact. 
The W nucleus has a deconvolved size of  
$0\farcs3 \times 0\farcs2$ (120 $\times$ 70 pc) 
and the E nucleus is unresolved by our 0\farcs5 beam (i.e. $\leq$ 0\farcs2) 
The equivalent blackbody temperatures 
for a disk of 0\farcs3 diameter are 29 K and 51 K for
E and W nuclei, respectively. 
Since the emission probably 
does not fill the maximum source area, 
the true brightness temperatures must be higher than these values.
A single component fit to the overall far infrared SED of Arp 220
yields a dust temperature of 42 K; however, this low temperature is difficult 
to reconcile with
the minimum brightness temperatures estimated above and the requirement that the dust be optically thin at 1 mm.
Thus, the overall spectral energy distribution
should be modeled more realistically with a range of dust temperatures
(mostly higher than 42 K) and probably an emissivity law steeper than $\nu^{1.3}$. The total dust mass of Arp 220 estimated assuming a single, optically thin component is $\sim 5 \times 10^{7} \Msol$ (Scoville \etal\ 1991). This
is probably an underestimate if multi-component models with moderate opacities
are more realistic.

\section{Molecular Gas Distribution and Kinematics}
	The CO emission peaks at the two continuum nuclei 
but is much more extended than the continuum emission. 
The total extent of CO in Fig.~\ref{total}d, in which most of 
single-dish flux is recovered, is about 2 kpc. 
The overall northeast--southwest velocity gradient seen in this larger scale feature suggests that the gas is rotating as a disk, as modeled by SYB.
The position angle of the velocity gradient, i.e., the major axis
of the disk, is measured to be $\sim 25 \arcdeg$.
It is perpendicular to the symmetric axis (P.A. $\sim 105$\arcdeg) of 
the bipolar \Halpha\ emission of Arp 220, 
which is in the shape of `double-bubble' of total extent 
$\sim 24 \times 10$ kpc
with one bubble on each side of the nucleus (Heckman, Armus, \& Miley 1987).
The bubbles have an expansion velocity of a few 100 \kms\ 
and blueshifted(redshifted) on the northwest(southeast) of the nucleus
(Heckman, Armus, \& Miley 1990).
If the kpc-scale molecular disk collimated the outflow of ionized gas, 
which was attributed to a starburst-driven superwind 
(Heckman \etal\ 1987, 1990),
then the orientation of the disk is such that the near side of
the disk is to the southeast.

About 30 \% of the total CO emission is intimately associated with the 
two nuclei as revealed most clearly in the U-array images.
The continuum and CO peaks in the U-array map coincide within 0\farcs05.
The systemic velocities of the two CO peaks (Table 1) agree 
well with those measured at the two NIR nuclei 
with the Br$\gamma$ line (Larkin \etal\ 1995).
The presence of two compact disk components associated with each of the double nuclei
was suggested by SYB (see their Fig. 11). Downes \& Solomon (1998)
clearly demonstrated these peaks but did not interpret them 
as disks (see below). One of the two models proposed by SYB assumed that
the gas distribution was smooth and axisymmetric.
These assumptions are not valid in the central 
part of the nuclear disk, i.e., within a diameter equal to the 
separation of the double nuclei.
Instead, their alternative model with peaks on each of the
double nuclei is more appropriate based on our higher resolution maps.
It was also noted by SYB that the CO emission profile on each nucleus 
was `double horned' with line width $\sim 250$ \kms, suggestive of disks rotating within each nucleus.
The two rotating gas disks were also inferred from
single-dish profiles of HCN and HCO+ lines (Taniguchi \& Shioya 1998).
The rotating mini-disk model now appears favored 
by our new data as discussed below.

\begin{figure}
 \vspace{14cm}
 \includegraphics{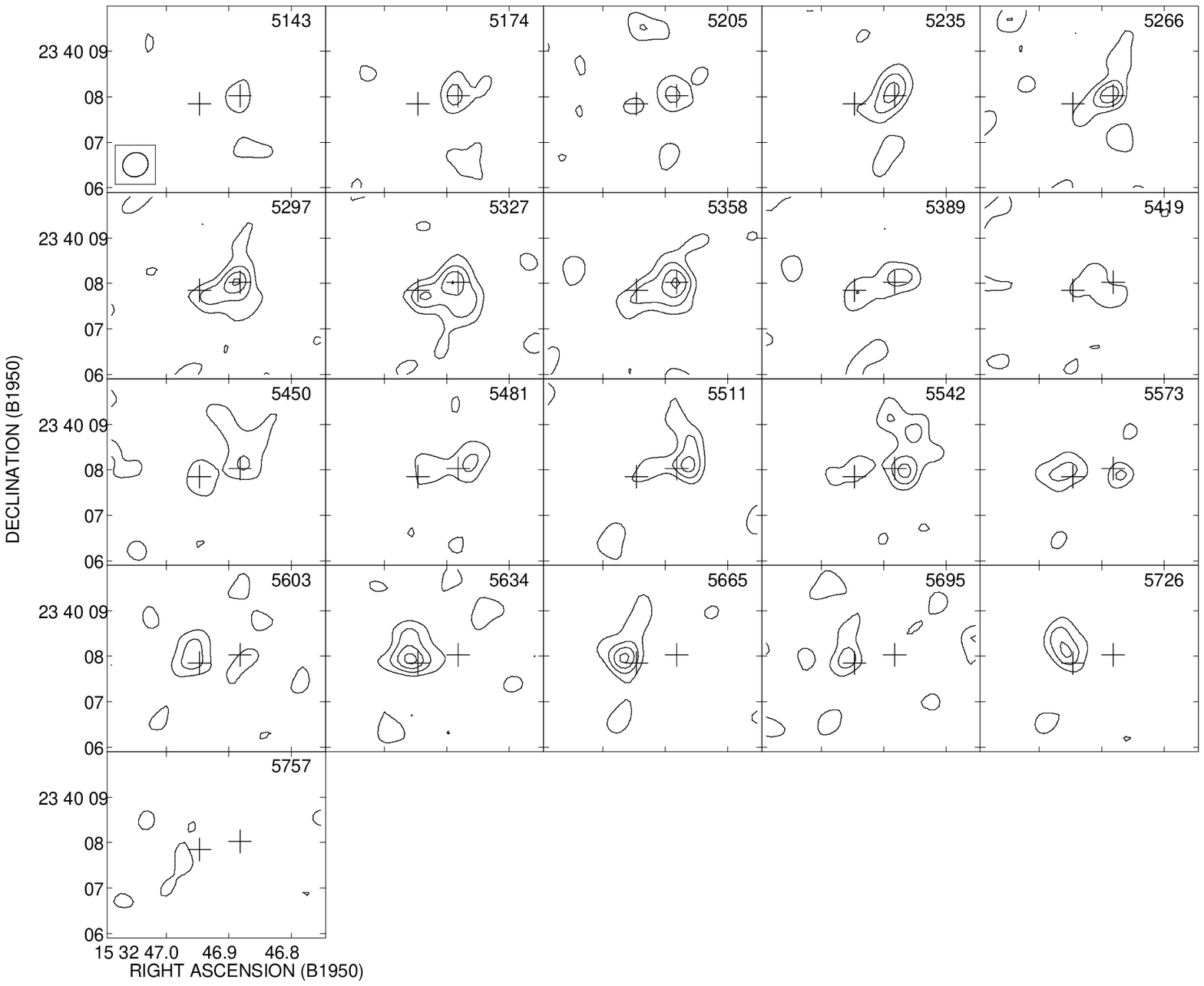}
 \vspace{-4cm}
 \caption{\footnotesize 
CO(2--1) continuum-subtracted channel maps from the U-array data. 
Contours are at 2, 3, 4, ..., 8 $\times$ 46 mJy \beam\ ($1\sigma$).
Crosses show the positions of the continuum peaks.
Velocities in \kms\ are shown in the northwest corner of all pannels.
The synthesized beam 
($0\farcs57 \times 0\farcs52$ FWHM, P.A.$= -60\arcdeg$)
is shown in the first pannel.
 \label{channel}}
\end{figure}

\begin{figure}
 \vspace{14cm}
 \includegraphics{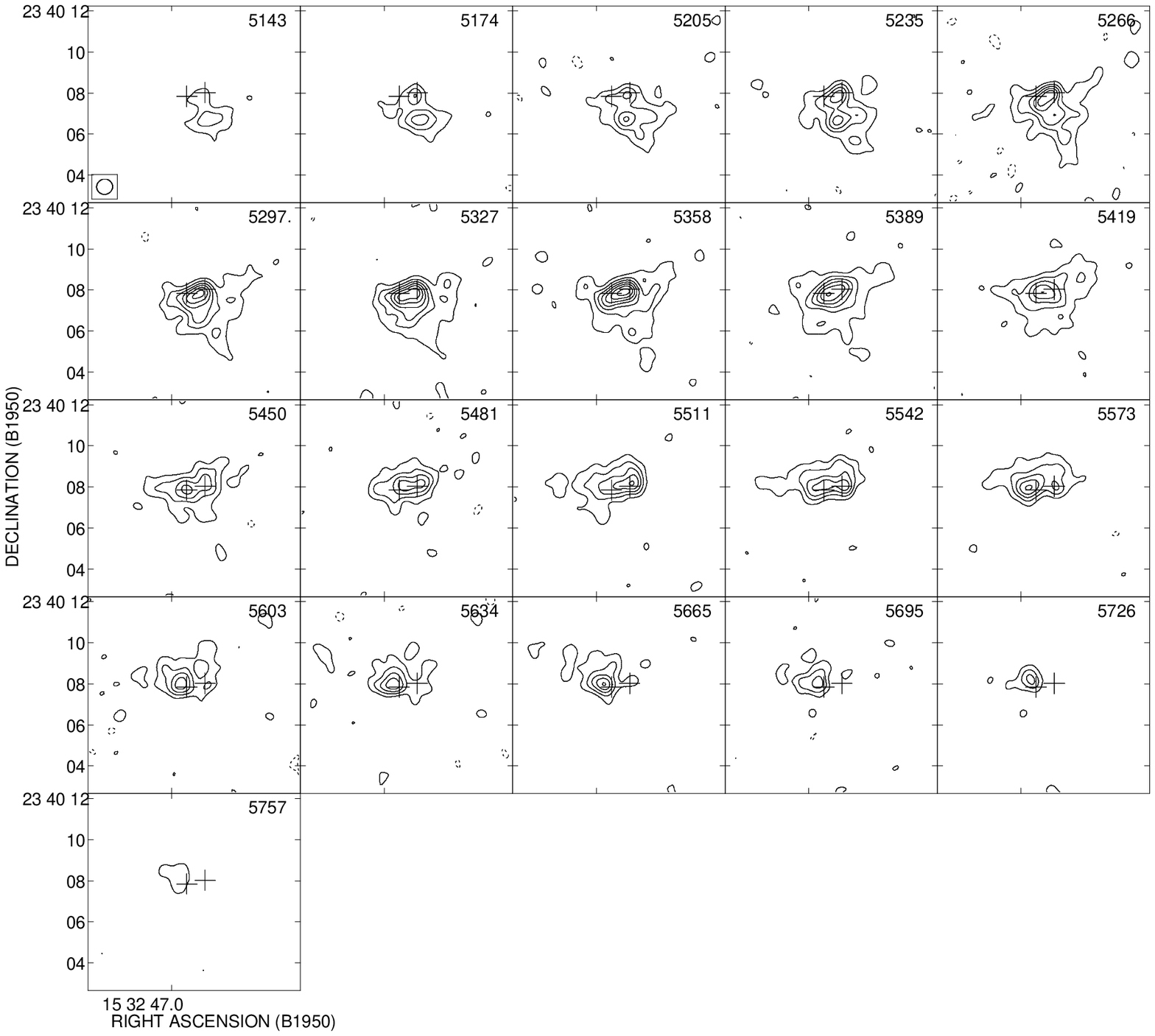}
 \vspace{-4cm}
 \caption{\footnotesize 
CO(2--1) continuum-subtracted channel maps from the LHU-array data. 
Contours are in steps of 90 mJy \beam\ ($3\sigma$).
Crosses show the positions of the continuum peaks.
Velocities in \kms\ are shown in the northwest corner of all pannels.
The synthesized beam 
($0\farcs78 \times 0\farcs74$ FWHM, P.A.$= -80\arcdeg$)
is shown in the first pannel.
 \label{chanLHU}}
\end{figure}

\begin{figure}[t]
{\hfill\epsfxsize=15cm\epsfbox{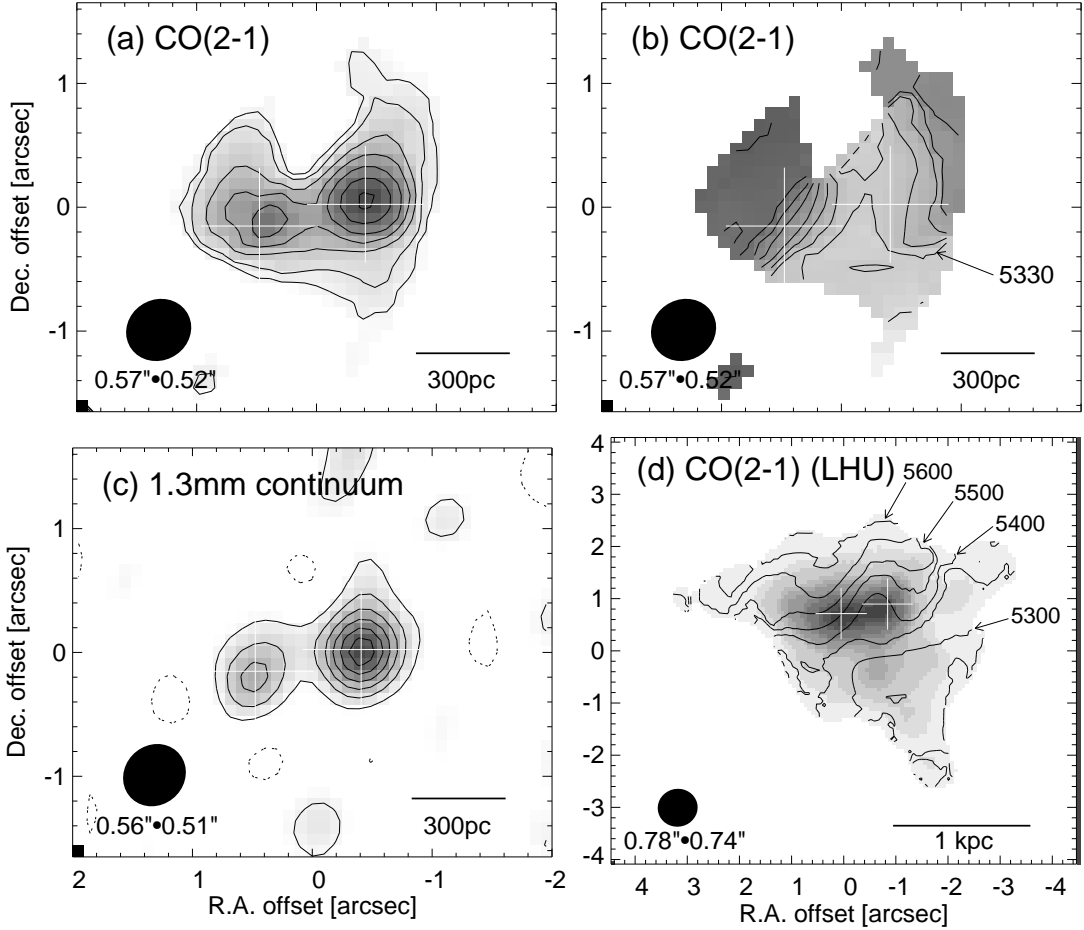}\hfill}
 \caption{\footnotesize 
Central region of Arp 220. 
Crosses in each panel indicate the 1.3 mm continuum positions of the nuclei.
(a) Continuum-subtracted CO(2--1) emission integrated over 645 \kms. 
Contours are at $8.5\times [1,2,4,6,8,10,12]$ Jy \beam\ \kms. 
(b) Mean velocity of CO(2--1) with contour intervals of 50 \kms. 
(c) 1.3 mm continuum with contour steps of 18.5 mJy \beam\ ($2\sigma$) and negative contours dashed.
(d) CO(2--1) integrated intensity map and isovelocity contours (50 \kms\ increments)
made from L, H, and U configuration data. 
The peak integrated intensity is 193 Jy \beam\ \kms.
 \label{total}}
\end{figure}

\begin{figure}[t]
{\hfill\epsfxsize=10cm\epsfbox{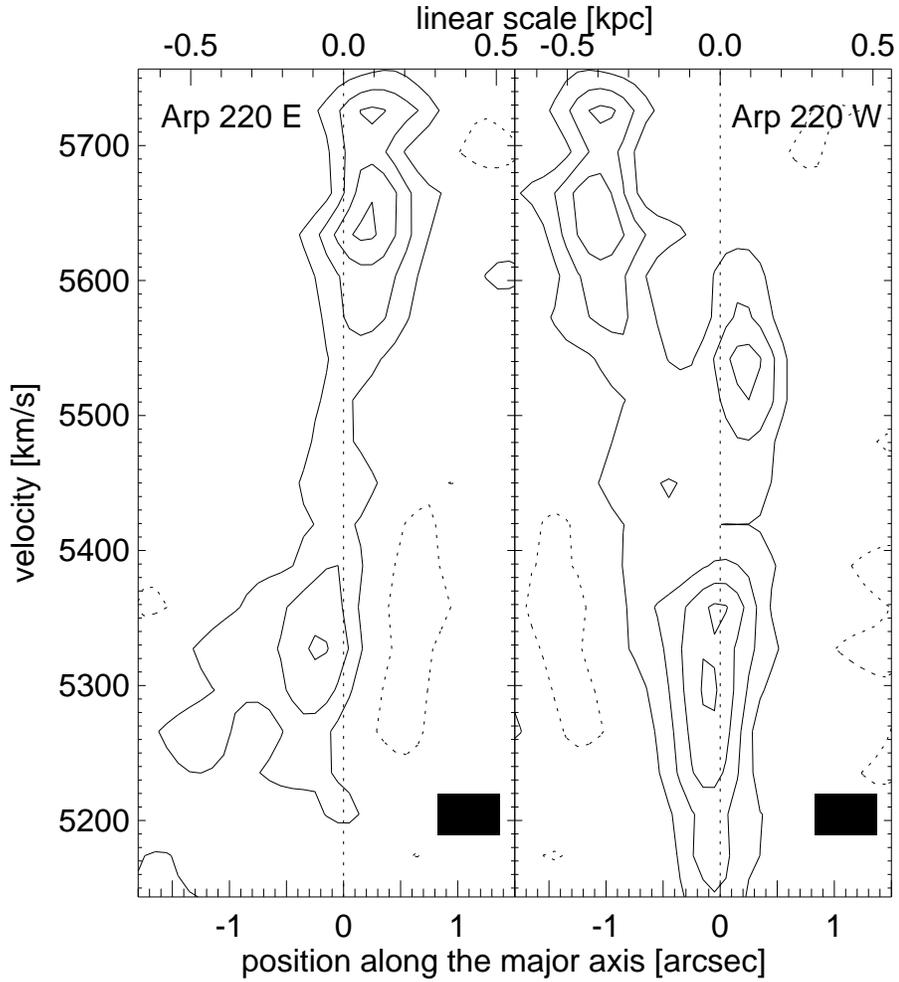}\hfill}
 \vspace{1cm}
 \caption{\footnotesize 
  CO(2--1) position-velocity maps along the major axes
  of the nuclear disks. 
  Contours are in steps of 46 mJy \beam\ \kms\ ($2\sigma$) without
  zero contours.
  The filled box in the lower-right corner of each pannel shows the
  spatial and velocity resolution.
  At $-1''$ in the P-V map of Arp 220 W, the emission is from 
  Arp 220 E, not W.
 \label{pv}}
\end{figure}

A steep velocity gradient is clearly seen across each nucleus 
in the U-array velocity map (Fig.~\ref{total}b), 
channel maps (Fig.~\ref{channel}), and also in the IRAM data (DS).
In the CO channel maps, the emission centroid continuously migrates from 
east of the cross (continuum position of Arp 220W) at lower velocities 
to west of the cross at high velocities. 
A similar shift of the emission centroid from southwest to 
northeast of the nucleus is also seen in Arp 220E; however,
it is apparent that the directions of the velocity gradients on E and W 
are not aligned. 
Moreover, the velocity gradient in Arp 220W at P.A.=263\arcdeg\ is skewed with 
respect to that in the outer disk (P.A.=25\arcdeg, Table~1).
The velocity shifts are about 500 \kms\ within 0\farcs3 (110 pc) 
of the nuclei as seen in the position-velocity maps (Fig.~\ref{pv})
as well as the channel maps (Fig.~\ref{channel}).
(Note that the mean velocities shown in Fig.~\ref{total}b average over the synthesized beamwidth which includes emission components
from both nuclei as well as the larger, P.A.= 25\arcdeg\ disk. The mean velocities therefore grossly underestimate the 
nuclear velocity gradients. In addition, the actual rotation curve should be determined
from the terminal or maximum velocities on each line of sight rather than the mean. The smaller line widths found in SYB are due to the
loss of low-intensity line wings in the clean component data and the larger 1\arcsec\ beam).

There have been several cm-wave interferometric studies of 
the velocity field in Arp 220 with somewhat contradictory results. 
Baan \& Haschick (1995) found a north-south velocity 
gradient within the west nucleus from VLA observations of 
formaldehyde (H$_{2}$CO) maser emission, 
while little velocity gradient was seen in the east nucleus.
The velocity gradient within the west nucleus is about 
$\Delta V = 125$ \kms\ in 1\arcsec\ in their mean velocity map.
The true velocity gradient could be larger if measured in a
position-velocity map along the line of nodes, 
though the map was not presented.
The velocity gradient at the west nucleus in the formaldehyde emission is
nearly perpendicular to that of CO. 
VLBI maps of the OH megamaser emission by Lonsdale \etal\ (1998) 
show two components of a few 10 pc extent in each nucleus.
The two in the west nucleus have $\sim 100$ \kms\ linewidths respectively
and are separated 0\farcs3 (110 pc) in the north-south direction, 
but no velocity gradient is seen between the two components.
The two in the east nucleus are separated about 50 pc 
in P.A. $\approx 30\arcdeg$, roughly in accord with
the major axis of the east CO disk (50\arcdeg).
The northeast component is redshifted by about 100 \kms\ relative
to the southwest component, also consistent with the CO velocity gradient.
The two components in each nucleus, however, have 
strikingly different morphologies,
i.e., one is linear with a bright broad-line knot in 
the middle and the other is in amorphous shape.
Therefore the two components in each nucleus do not necessarily
indicate the elongation and rotation of a gas disk.
Lastly, recent mapping of the HI absorption
by Mundell, Ferruit \& Pedlar (1998) 
shows a roughly NE--SW velocity gradient in the east nucleus, 
similar to that found in CO, while velocity structure in the
west nucleus is too complicated to be simply modeled with a 
velocity gradient.
The complicated velocity structure in the western nucleus
could reflect warping of the disk on different scales 
(since the size of the observed regions ranges by more than
 an order of magnitude) 
or the selection of favorable gain paths
for maser amplification of the nuclear continuum source(s).

\section{Nuclear Gas Disks}
\begin{figure}[t]
{\hfill\epsfxsize=10cm\epsfbox{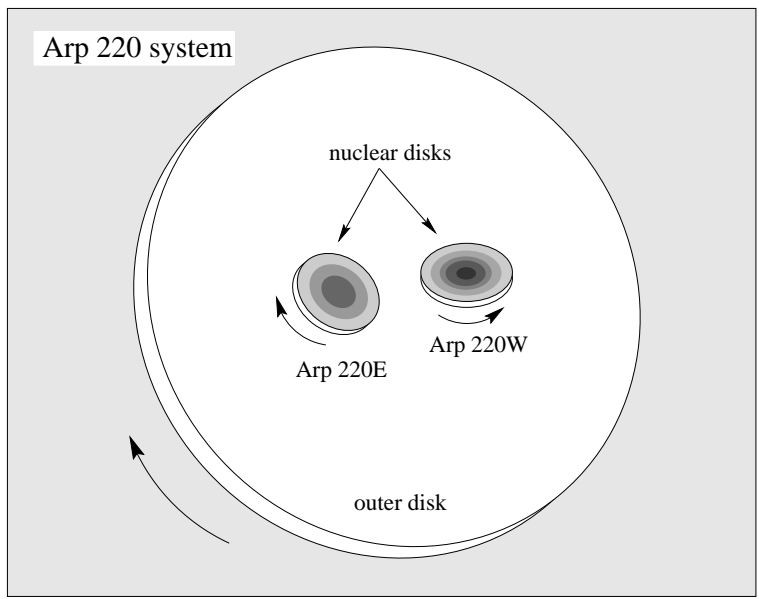}\hfill}
 \caption{\footnotesize 
Schematic illustration of the Arp 220 disks.
The two nuclear disks have radii $\sim 100$ pc and misaligned spin axes, probably counterrotating with respect to each other.
The dynamical mass within 100 pc radius of each nuclear disk is about
$2\times 10^{9} \sin^{-2} i$ \Msol\ where $i$ is the inclination of
the disk. 
Gas masses in the nuclear disks are estimated to be $\sim 10^{9}$ \Msol\
for each.
Each of the nuclei contains a gas disk, young stars formed in the disk,
and presumably an old stellar population from the nucleus of the merger progenitor.
The stellar component likely has a larger scale height than molecular gas,
and thus the gas disk is embedded in the stellar nucleus.
Most of the far-IR luminosity of Arp 220 is from the central 100 pc diameter
of the nuclear disks.
The outer disk has a radius of $\sim 1$ kpc, 
formed from the gas from the progenitor galaxies and now
rotating around the dynamical center of the merger.
It is likely the host of a previous starburst of $\sim 10^{8}$ yrs ago 
that created the \Halpha\ bubbles $\sim$ 10 kpc in size. 
 \label{illust}}
\end{figure}

The steep velocity gradients as well as the misalignments of the 
CO nuclear disk structures exclude the
possibility that the gas is dynamically coherent with the larger disk.
Instead, our data suggest that each nucleus has a separate, rotating 
molecular gas disk of diameter about 200 pc (0\farcs6).
The directions of the velocity gradients imply that the rotation
axes of these two nuclear disks are misaligned.
The western nuclear disk with the major axis in east--west direction
was inferred from the crescent 
morphology of the 2 $\mu$m emission in HST NICMOS images (Scoville \etal\ 1998).
The sharp falloff of light to the south of the western nucleus in the
NICMOS images suggests that the near side (the side of higher extinction) 
of the west disk is to the south.
The NICMOS imagery also resolved the eastern nucleus into two peaks separated
by 0\farcs4 in the north-south direction with the southern peak being more highly reddened. 
It is therefore likely that the near side of the eastern nuclear disk
is also to the south. For these orientations (with the near sides of
both disks on the south), the two nuclear disks must be counterrotating.
The suggested configuration of the gas disks in Arp 220 is illustrated
in Fig.~\ref{illust}.

Downes \& Solomon (1998) have interpreted the E and W peaks as starbursting gas clumps formed from gravitational
instabilities in the larger gas disk. They interpret the velocity gradients across
each nucleus as galactic bar streaming motions or starburst-driven molecular outflows.
However, the magnitude of the observed velocity gradients ($\sim$ 500 \kms) 
and their different position angles across
each CO peak (nearly coincident with the dust continuum, non-thermal radio continuum, and  near infrared peaks) is strong evidence that 
these peaks are high mass concentrations (therefore probably real galactic nuclei as opposed to starburst luminosity peaks).
The velocity range at each nucleus is in fact comparable with the
overall linewidth of the CO over the entire nucleus of Arp 220 
and thus the velocity gradients can not be accounted for as minor
perturbations (e.g., streaming) in a bar. Similarly, outflow winds
at several hundred \kms\ would have difficulty remaining molecular and
their expected position angle should be perpendicular to the central disk.
(Although the position angle of the velocity gradient at the west nucleus
is almost perpendicular to the larger disk and aligned to the \Halpha\
`double-bubble', the direction of the velocity gradient is {\em opposite} 
to that of the \Halpha\ outflow.)
Lastly, it is unclear how a starburst-generated outflow could 
contain a significant fraction of the same dense gas which sustains the starburst.  

Whether the mass concentrations seen at the two peaks contain
a significant contribution of mass from the pre-merger galactic nuclei 
or is almost entirely gas plus young stars is difficult to answer on the basis of existing {\it observational} data. 
The absence of evidence of old stars in the optical/IR 
spectrum of Arp 220 is certainly not evidence of their absence given 
the fact that the younger, massive stars are much brighter. 
Downes \& Solomon (1998) argued against a significant mass from the 
old stellar population on the basis of 
the mass budget --- when they add their estimates of the mass in 
the molecular gas and the young stars required to account 
for the observed luminosity, this sum approaches the dynamical mass 
and hence leaves little room for old stars. However, this does not seem a strong basis to 
rule out an older stellar population since the
dynamical mass is really a lower limit (due to the uncertain inclination); 
the molecular mass estimate is clearly uncertain given the 
extreme excitation conditions; and lastly, the derived young stellar mass
depends sensitively on the starburst IMF and the duration of the starbursts
(cf. SYB and Leitherer \& Heckman 1995).

\subsection{Comparison of Dust and CO Emission}
	The marked difference in the extent of CO and continuum emission
is real, not an artifact of missing flux in the continuum map.
The LHU-array CO map made with all channels
combined before FFT (i.e., processed in the same way as continuum maps) 
still shows the CO emission extended well outside the double nuclei.
The striking difference between the CO and dust spatial extents
could be due to : line-of-sight blocking of the CO emission (to suppress that from the 
nuclear regions); different dependences of the CO and dust emissivities 
on the physical conditions (principally temperature $T$ and column density); 
or a reduction in the CO abundance relative to
the dust in the ISM of the nuclear clusters. The idea of overlapping 
CO emission clouds was proposed for 
the larger gas disk of Arp 220 by Downes, Solomon, \& Radford (1993)
and this idea could be extended to postulate that there
is even more severe overlap in the compact nuclear disks than in the 
larger disk. If the gas cloud density in each nuclear disk is so high that the clouds overlap both on the sky and in line-of-sight velocity, then the
intensity of the optically thick CO will reflect the gas temperature at the disk surface in the nearest cloud at each velocity. [To be a significant effect, it is required that the stars in the disk dominate the potential of 
the gas; otherwise, if the gas actually contributes most of the mass,  
increases in the mass surface density of the disk cause comparable increases in the CO
line width and hence yield increased velocity-integrated emission.]

Alternatively, if the dust and CO emission had different dependences on the
physical conditions which are likely to be functions of radius, then the dust continuum could be more centrally peaked than the CO. The optically thin dust emission is proportional to the product of the dust temperature and column density while the CO surface brightness (assuming the CO is optically
thick) will vary as product of : the CO column density; a factor proportional to the ratio $T_{\rm ex}/\langle n_{\rm H_2} \rangle^{1/2}$,
where $T_{\rm ex}$ is the excitation temperature 
and $\langle n_{\rm H_2} \rangle$ is the mean H$_{2}$ density; 
and a geometry-dependent factor
which reflects the importance of external gravitational 
potentials and pressures in the emission line width and cloud
binding (cf. Downes \etal\ 1993, Bryant \& Scoville 1996).
Thus, even with a constant ratio of the CO and dust abundances, 
a dense ISM like that in the nuclear disks of Arp 220 might naturally
have relatively strong, optically thin dust emission 
if temerature and density increases at small radii without
increasing $T_{\rm ex}/\langle n_{\rm H_2} \rangle^{1/2}$.
  
Lastly, a low CO-to-continuum ratio near the
nuclei could occur if the CO is optically thin due to high temperatures (DS)
or the gas is depleted relative to the dust in the vicinity 
of the nuclei.
However, since the observed CO brightness temperatures ($T_{\rm b} \sim 38$ K) 
are comparable to
the mean dust temperature (42 K) derived from fitting the far infrared SED,
low opacity in the CO lines would necessitate that the gas is much hotter and not in thermal equilibrium with the dust (despite the general 
expectation that they should be in thermal equilibrium at densities 
$\geq 10^{4-5}$ cm$^{-3}$). In addition, the CO 2-1/1-0 line ratios 
indicate that the lower CO transitions cannot be optically thin. 
It is therefore unlikely that bulk of the CO emission from the nuclear gas
disks is optically thin. 

It is useful to review the necessity of continuum subtraction
for the situation where CO emission is optically thick and the clouds are
overlapping. Subtraction of the continuum from the emission line channels is 
appropriate if, at each velocity,
the optically thick CO covers only a small fraction of the projected area
of the continuum source. On the other hand, if the continuum source is embedded in and totally covered by
the optically thick CO emission at {\it each} velocity, then continuum subtraction 
should be done only outside the velocity range of the CO line. The CO fluxes of the nuclear disks (Table 1) were measured 
from continuum-subtracted data 
and thus underestimate the true CO flux and brightness temperature
in the latter case. Without continuum subtraction, the total CO flux and the peak brightness temperature of the nuclear gas disks are 43 K and 150 Jy \kms\ for Arp 220E
and 48 K and 260 Jy \kms\ for Arp 220W --- these flux values are about 30 \% larger than those given in Table 1.
The real situation is probably between these extreme cases, i.e.,
the continuum emission is partially covered by optically thick gas but closer
to the low covering situation since only a small
fraction of the full line width is represented along each line of sight to the 
continuum source. In any case, the continuum subtraction does not affect the velocity gradients
seen in the nuclear disks since similar gradients are seen in both
the continuum subtracted and non-subtracted line maps.

\subsection{Gas and Dynamical Masses of the Nuclear Disks\label{s.masses}}
The dynamical mass in each nuclear disk within 100 pc radius
is $\sim$2 $\times 
10^9\sin^{-2}{i}\; \Msol$, where $i$ is the inclination of the disk
(see Table 1). 
A thin disk model with parameters of 
$V_{\rm rot} \sin i \sim 265 \pm 10$ \kms\
at $r_{\rm peak} \sim 0\farcs25$  
produced a reasonably good fit to the position--velocity diagram
of Arp 220E.
The presence of the two nuclear disks, their misaligned rotational axes,
and their large masses strongly implies that the two
peaks in millimeter, radio, and NIR are major mass concentrations (i.e., 
galactic nuclei), 
and that Arp 220 is indeed a merger.
The two nuclear disks are embedded in an outer gas disk of 
kiloparsec size.
The dynamical mass within the orbit of the two nuclei (i.e., $r < 250$ pc) 
was estimated to be $5.4 \times 10^{9} \Msol$ by SYB.
Thus more than half of the mass in this region belongs to the
two nuclei, though the dynamical masses have large uncertainties 
due to the uncertain inclinations.

Estimation of the molecular gas mass from the observed CO emission
in the nuclear disks of Arp 220 is extremely uncertain. 
The physical conditions of the gas are undoubtedly very different 
from those of the Galactic GMCs where the `standard' Galactic CO-to-H$_2$ conversion factor is derived. Moreover, it has been pointed out by 
Downes \etal\ (1993) that if the large-scale 
gravitational field of the galactic nucleus (rather than the 
self-gravity of individual clouds) dominates the structure of the 
nuclear CO emission regions, the escape probability of 
optically thick CO line photons will be increased (due to the 
larger velocity width), thus reducing the CO-to-H$_2$ conversion factor.
Scoville, Yun \& Bryant (1997) estimated an average CO-to-H$_2$ conversion factor of 0.45 times the Galactic value for the larger molecular 
disk in Arp 220. Their method was based on fitting the observed line profiles
to an axisymmetric disk model and assuming that the enclosed molecular 
gas mass determined the disk rotation velocity as a function of radius. 
This {\it average} conversion factor is probably inapplicable to the very
compact nuclear 
gas disks because of the more extreme physical conditions and the likely 
contribution of stars to the mass within the nuclei.

Here, we follow the discussion in the Appendix of Bryant \& Scoville (1996)
to place constraints on the molecular gas mass.
A firm lower limit of $1\times 10^{8} \Msol$ for Arp 220W 
(and 2/3 as much for E), 
is obtained if the CO is optically thin 
and in LTE at 40 K with a CO-to-H$_2$ abundance of $10^{-4}$.
For optically thick CO emission from self-gravitating GMCs, the higher gas temperature in Arp 220 can be compensated to some extent
by the higher gas density since the conversion factor scales as the ratio
$\langle n_{\rm H_{2}} \rangle^{1/2}/T_{\rm b}$. 
If the $T_{\rm b}$ is 50 and 10 K
in the Arp 220 and Galactic GMCs respectively and the mean volume 
densities are 10$^{4.5}$ and 300 cm$^{-3}$, then the conversion factor is 
changed ({\it actually increased}) by only a factor of 2 due to the
compensating changes in temperature and density.
The higher gas density is implied by the observations of
HCN, a tracer of dense gas. The HCN-to-CO line ratio in Arp 220 requires
a mean density exceeding $10^4$ cm$^{-3}$ (Solomon, Downes, \& Radford 1992).
In addition, this density is consistent with the assumption that the mass derived below is distributed in a disk of 100 pc radius and thickness $\sim$
30 pc.
Modeling the effects of the 
galactic nucleus gravitational potential as a disk, there are two possibilities : an entirely self-gravitating gas disk and a gas disk embedded
in a potential predominantly due to stars.
If the gas disk is geometrically thin (with inclination $i$), smooth, and self-gravitating, 
then the gas mass in W is $1.3\times 10^{9} \sec i$ \Msol\ for gas with
density and brightness temperature having a ratio of 
$\langle n_{\rm H_{2}} \rangle^{1/2}/T_{\rm b}$ similar to Galactic GMCs.
For an inclination of 45\arcdeg, the conversion factor in the disk is then 0.4 times
the Galactic value and the gas masses in E and W are 1--2 $\times 10^{9}$ \Msol,
which agrees within a factor of 2 to the values derived by DS using the
similar method of Downes \etal\ (1993).
If the gas disk is not entirely self-gravitating but bound partly by the
stellar (or central black hole) gravity then the gas mass in W is
$1.6\times 10^{9} f_{\rm p}^{1/2} \sec i$ \Msol, 
where $f_{\rm p}$ ($\ll 1$) 
is a parameter describing the relative contribution of gas and stars to
the kinematics of the disk.
[Specifically, $f_{\rm p}$ is the ratio of gaseous to stellar mass densities
at the midplane of the disk (see Bryant \& Scoville for details). 
Our observations do not constrain this parameter and thus the gas mass
can not be evaluated in this case.]
A firm upper limit of gas mass is $M_{\rm gas} \leq M_{\rm dyn}$.

\section{Merger Evolution and Luminosity Source}
Numerical simulations of mergers (e.g., Barnes \& Hernquist 1991) 
have shown that the gas concentrates at the center of
each galaxy in the early phases of interaction, 
with some delay after the first encounter when 
progenitors have massive bulges
(Mihos \& Hernquist 1994).
Later, the gas condensations merge and
form a single gas disk as they approach each other.
Our results are largely consistent with this scenario since
the larger gas disk extending to 1 kpc radius has 70 \% of CO luminosity.
However, it is clear from the results presented here that each
nucleus also has its own smaller, denser disk, surviving
until the merging is nearly complete. These compact, nuclear
disks are probably remnants of the gas which concentrated within the
nuclei when they were much further apart. Presumably any gas 
concentrated within the deep potentials of the original nuclei
where the escape velocity is $\geq 300~\kms$ would be trapped until
it is either exhausted in star formation or AGN accretion or
ejected in a more diffuse wind generated by the starburst. The latter 
would probably be inhibited as a major mechanism of mass-loss until the 
disk has evolved to much lower column density than is presently the case. 
As long as there is enough
cold, dense gas in the disk to dissipate and radiate the wind kinetic energy,
the outflows generated by starburst activity embedded within the
disks can not rip much molecular gas off the disks.

Our observations do not give a conclusive answer regarding the origin
of the large FIR luminosity of Arp 220, since the dust temperature
is consistent with the dominant heating source 
being either a starburst or an AGN.
However, our results do give the following spatial restriction on the energy source.
The fact that the 1 mm continuum fits on the FIR tail of the 
overall spectral energy distribution (containing most of the bolometric luminosity)
suggests that spatial distribution of the energy source(s)
is similar to that of the 1 mm continuum i.e., dominated by two components 
with the western source being probably twice as luminous. 
The size of the two luminosity sources must also be $\lesssim 100$ pc.
It is possible for additional components to contribute to the
bolometric luminosity (without contributing 1 mm flux)
only if the temperatures of these components are much higher than for the sources at the nuclei. Since it seems unlikely that dust in the outer gas disk should have  higher 
temperatures than that in the nuclei, we conclude that only a small 
portion of the luminosity originates from the 1 kpc disk. 
[We therefore differ from DS who attribute half of the luminosity
to the outer molecular disk but as noted earlier, they claim a 
larger fraction of the 1 mm continuum in an extended component.]

The suggested distribution of luminosity is consistent with either
a double starburst or double AGN for the energy source. 
It excludes a single, dominant AGN or a starburst in the gas concentration toward 
the dynamical center, between the two nuclei. 
Since the energy sources are in both nuclei, the elevated luminosity of Arp 220
was probably triggered by gas condensation within the nuclei of 
the progenitor galaxies.
The upper-limit of 100 pc is comparable to the 
spatial extent of the unresolved VLBI sources, probably radio supernovae, 
around the west nucleus (Smith \etal\ 1998).
If the bulk of the luminosity of Arp 220 arises from starbursts 
(e.g., Sturm \etal\ 1996; Lutz \etal\ 1996; 
Genzel \etal\ 1998, and references therein),
the starbursts are embedded in the two nuclear gas disks and their
radiation would be highly extincted. The average
extinction perpendicular to the disks is $A_{V} \sim 1000$ mag
based on the mean gas and dust surface densities. 
Armus \etal\ (1995) suggested based on NIR photometry and spectroscopy 
(CO absorption and Br$\gamma$) that there are young (several $10^{6}$ yr old) 
starbursts at the two nuclei surrounded by an old ($\sim 10^{8}$ yr old) 
starburst region of $\sim$ 5\arcsec\ (2 kpc) diameter. 
In their accounting, these starbursts contribute less than 10 \% to
the bolometric luminosity of Arp 220; however, they adopted  
10 mag of visual extinction in a foreground screen whereas the
ISO observations indicate probably at least 45 mag for a screen 
model or A$_{V}\sim 1000 $ mag for a mixed source/dust model (Lutz \etal\ 1996).
The older starburst appears to have occured in the outer kiloparsec-size
molecular disk and probably caused the large \Halpha\ `double-bubble' 
whose dynamical age is $\sim 10^{8}$ yrs (Heckman \etal\ 1990).
The current starbursts in the nuclear disks suffer from large extinction
and thus could contribute more to the bolometric luminosity than
the above estimate from NIR observations.

Another restriction for the luminosity source is the average 
luminosity-to-mass ratio $L_{\rm IR}/M_{\rm dyn}$ in the nuclear disks. 
It is as high as 200 \Lsol/\Msol\ for 
nominal inclinations of 45\arcdeg.
Such a high ratio can not be attained unless the initial mass 
function of the starbursts is biased toward high masses 
(e.g., truncated below a few $\Msol$) 
or there are
additional energy sources such as heavily obscured AGNs 
(SYB; Shier, Rieke, \& Rieke 1996).
The large mass of gas in the nuclear disks can fuel
both starbursts and AGNs for $\geq 10^{8}$ yrs whereas 
the rotational kinetic energy of 
the nuclear disks ($\approx$10$^{50}$ J for each) 
can sustain the FIR luminosity for just $10^4$ yrs.
Assuming that the nucleus is not viewed close to 
the polar axis, the high opacities of the nuclear ISM can easily 
absorb the direct emission of AGNs (if any exist) from moderately high energy  
X-ray and to mid-IR wavelengths.
The lack of a compact, high brightness temperature nuclear radio source 
also does not rule out an AGN 
since not all AGNs have strong radio emission (Smith \etal\ 1998).
Lonsdale \etal\ pointed out that the linear morphology and broad
linewidth of one of the two OH megamaser components in each nucleus
could be due to twin jets and a molecular torus of an AGN, 
though they did not exculde the possibility of starburst-driven
shocks for the excitation mechanism of the OH maser.

\section{Conclusions}
	Our 0\farcs5 resolution observations of CO and dust continuum 
provide the following picture 
of the central region of Arp 220:
There are two counterrotating gas (and dust) disks of $\sim$100 pc radius
within each of the double nuclei. This counterrotation may reflect
the geometry of the progenitor galactic disks in their 
mutual encounter. If so, it would facilitate the eventual 
merging of the galaxies due to the more effective 
cancellation of angular momentum and dissipation of kinetic energy
in such encounters.    
These nuclear disks are embedded in a kiloparsec-size gas disk, which 
rotates around the dynamical center of the merger.
The gas and dust in the nuclear disks is hot ($\geq 40$ K), dense 
and probably covers at least half of the nuclear disk surface area.
The dynamical masses within 100 pc radius of each nucleus are 
$\gtrsim 2 \times 10^{9}$ \Msol. 
The FIR luminosity is generated predominantly within the central 100 pc of 
the two nuclei with a high luminosity-to-mass ratio.

\vspace{5mm}

	We are grateful to the OVRO staff for their hard work that lead
to the smooth startup of the new configuration.
We thank Dr. Carole Mundell for kindly communicating the results of
HI observations prior to publication.
The OVRO mm-array is funded by NSF grant AST 96-13717 
and the K. T. \& E. L. Norris Foundation.
KS is supported by JSPS fellowship.

\clearpage
\begin{deluxetable}{llccc}
\tablecaption{Parameters of Arp 220}
\tablewidth{0pt}
\tablehead{
 \colhead{Parameter } & 
 \colhead{unit} &
 \colhead{Arp 220 E} &
 \colhead{Arp 220 W} & 
 \colhead{Total} 
}
\startdata
position \ ($\Delta\alpha, \Delta\delta$)\tm{a} & ($''$)& $(0.89, -0.18)$ & $(0, 0)$ 	& \nodata  	\nl
$S_{\rm 1.3mm}$               	& (mJy)	 	& 66		& 142		& 208\tm{b}   \nl
deconvolved size at 1.3 mm      & (pc)			& unresolved	& $120\times 70$ (p.a. 170\arcdeg)  & \nodata \nl 		     $\Delta T_{\rm equiv, 1.3 mm}$  & (K)\tm{c}		& 29		& 51		& \nodata	\nl
$S_{\rm 3.0mm}$               	& (mJy)\tm{d}		& (9.5)		& (19)		& 34 		\nl
$S_{\rm CO(2-1)} $ (OVRO U-array) & (Jy \kms)		& 120		& 187   	& 307\tm{e}   	\nl
$\Delta T_{\rm b, CO(2-1)}^{\rm peak}$ & (K)\tm{f}  	& 38		& 37		& \nodata	\nl
$V_{\rm mean}$		      	& (\kms)\tm{g}		& 5523		& 5350  	& \nodata	\nl
P.A. of velocity gradient     	& (\arcdeg)\tm{h} 	& 52		& 263		& 25		\nl
$\Delta V(r\leq 0\farcs3)$ 	& (\kms)\tm{i}		& 540		& $>480$ 	& \nodata  	\nl
$M_{\rm dyn}\sin^2 i$           & (\Msol)\tm{j} 	& $1.9\times10^9$ & $>1.5\times10^9$ 	& $5.4\times10^{9}$ \nl	
$M_{\rm gas}$			& (\Msol)\tm{k}		& $\sim 10^9$	& $\sim 10^9$   & \nodata	\nl
\enddata
\tablenotetext{a}{
Arp 220 W is at
R.A.=15$^{\rm h}$32$^{\rm m}$46\fs88, Dec.=+23\arcdeg 40\arcmin 08\farcs0 (B1950) 
($\pm$ 0\farcs1). The continuum and line peaks coincide within 0\farcs05.
}
\tablenotetext{b}{
Single-dish flux scaled from 1.25 mm (Carico \etal\ 1992) assuming 
$\kappa_{\nu} \propto \nu^{1.3}$ is $213\pm10$ mJy.
}
\tablenotetext{c}{
Equivalent excess blackbody temperature for an emitting region of a 0\farcs3 diameter. 
}
\tablenotetext{d}{
The total flux density was measured in a 
1\farcs2 resolution map, which partially resolves the two nuclei.
Flux density of each nucleus is the sum of CLEAN components, giving
a lower limit.
}
\tablenotetext{e}{
Single-dish total flux measured by Radford \etal\ (1991) is 1038 Jy \kms.
}
\tablenotetext{f}{
Peak excess brightness temperature of CO(2--1) line in the channel maps. 
}
\tablenotetext{g}{
Heliocentric CO(2--1) velocity ($cz$) at the nucleus. 
}
\tablenotetext{h}{
Position angle of the velocity gradient at the nucleus. 
The last column is for the outer disk.
}
\tablenotetext{i}{
Line-of-sight velocity width. Our spectrometer does not fully cover Arp 220 W.}
\tablenotetext{j}{
The dynamical masses for E and W are those within a radius of 0\farcs3 (112 pc) without
inclination correction. The total dynamical mass is that within 
a radius of 250 pc corrected for an inclination of 45\arcdeg\ (SYB).
}
\tablenotetext{k}{
see text (\S \ref{s.masses}).
}
\end{deluxetable}

\clearpage

\clearpage

\clearpage

\clearpage

\clearpage


\clearpage

\begin{thebibliography}{}
\bibitem[Armus \etal 1995]{Armus95}
	Armus, L., Neugebauer, G., Soifer, B. T., Matthews, K. 1995, \aj, 110, 2610
\bibitem[Arp 1966]{Arp66}
	Arp, H. 1966, {\it Atlas of Peculiar Galaxies}, 
	(Pasadena: California Institute of Technology)
\bibitem[Baan \& Haschick 1995]{Baan95}
	Baan, W. A., \& Haschick, A. D. 1995, \apj, 454, 745
\bibitem[Barnes \& Hernquist 1991]{BH91}
	Barnes, J. E., \& Hernquist, L. E. 1991, \apj, 370, L65
\bibitem[Bryant \& Scoville 1996]{BS96}
	Bryant, P. M. \& Scoville, N. Z. 1996, \apj, 457, 678
\bibitem[Carico et al. 1992]{Carico92}
	Carico, D. P., Keen, J., Soifer, B. T., \& Neugebauer, G. 1992, \pasp, 104, 1086
\bibitem[Downes & Solomon 1998]{Downes98}
	Downes, D., \& Solomon, P. 1998, \apj, in press
\bibitem[Downes et al. 1993]{Downes93}
	Downes, D., Solomon, P., \& Radford, S. J. E. 1993, \apj, 414, L13
\bibitem[Genzel et al. 1998]{Genzel98}
	Genzel, R., Lutz, D., Sturm, E., Egami, E., Kunze, D., Moorwood, A. F. M., Rigopoulou, D., Spoon, H. W. W., Sternberg, A., Tacconi-Garman, L. E., Tacconi, L., and Thatte, N. 1998, \apj, 498, 579
\bibitem[Graham et al. 1990]{Graham90}
	Graham, J. R., Carico, D. P., Matthews, K., Neugebauer, G., Soifer, B. T., and Wilson, T. D. 1990, \apj, 354, L5
\bibitem[Heckman \etal\ 1987]{HAM87}
	Heckman, T. M., Armus, L., and Miley, G. K. 1987, \aj, 92, 276
\bibitem[Heckman \etal\ 1990]{HAM90}
	Heckman, T. M., Armus, L., and Miley, G. K. 1987, \apjs, 74, 833
\bibitem[Larkin 1995]{Larkin95}
	Larkin, J. E., Armus, L, Knop, R. A., Matthews, K, and Soifer, B. T.  1995, \apj, 452, 599
\bibitem[Leitherer \& Heckman 1997]{Leither97}
	Leitherer, C. \& Heckman, T. 1995, \apjs, 96, 9
\bibitem[Lonsdale \etal\ 1998]{Lonsdale98}
	Lonsdale, C. J., Diamond, P. J., Smith, H. E., and Londale, C. J. 1998, \apjl, 493, L13 
\bibitem[Lutz 1996]{Lutz96}
	Lutz, D., Gnezel, R., Sternberg, A., Netzer, H., Kunze, D., Rigopoulou, D, Sturm, E., Egami, E., Feuchtgruber, H., Moorwood, A. F. M., and de Graauw, Th. 1996, \aap, 315, L137
\bibitem[Mihos \& Hernquist 1994]{Mihos94}
	Mihos, J. C., \& Hernquist, L. 1994, \apj, 431, L9
\bibitem[Mundell 1998]{Mundel98}
	Mundell, C. G., Ferruit, P. \& Pedlar, A. 1998, in preparation
\bibitem[Norris 1985]{Norris85}
	Norris, R. P. 1985, \mnras, 216, 701
\bibitem[Radford et al. 1991]{Radford91}
	Radford, S. J. E., Solomon, P. M., \& Downes, D. 1991, \apj, 368, L15
\bibitem[Sanders et al. 1988]{Sanders88}
	Sanders, D. B., Soifer, B. T., Elias, J. H., Madore, B. F., Matthews, K., Neugebauer, G., and Scoville, N. Z. 1988, \apj, 325, 74
\bibitem[Sanders et al. 1991]{SSS91}
	Sanders, D. B. Scoville, N. Z., Soifer, B. T. 1991, \apj, 370, 158
\bibitem[Scoville et al. 1986]{Scoville86}
	Scoville, N. Z., Sanders, D. B., Sargent, A. I., Soifer, B. T., Scott, S. L., and Lo, K. Y. 1986, \apj, 311, L47
\bibitem[Scoville et al. 1991]{S4}
	Scoville, N. Z., Sargent, A. I., Sanders, D. B., \& Soifer, B. T. 1991, \apj, 366, L5 
\bibitem[Scoville et al. 1993]{Sco93}
	Scoville, N. Z., Carlstrom, J. E., Chandler, C. J., Phillips, J. A., 
	Scott, S. L., Tilanus, R. P. J., \& Wang, Z. 1993, \pasp, 105, 1482
\bibitem[Scoville et al. 1997]{SYB}
	Scoville, N. Z. Yun, M. S., \& Bryant, P. M. 1997, \apj, 484, 702 (SYB)
\bibitem[Scoville et al. 1998]{Scoville98}
	Scoville, N. Z., Evans, A. S., Dinshaw, N., Thompson, R., Rieke, M., Schneider, G., Low, F. J., Hines, D., Stobie, B., Becklin, E., and Epps, H. 1998, \apj, 492, L107
\bibitem[Shier et al. 1996]{Shier96}
	Shier, L. M., Rieke, M. J., \& Rieke, G. H. 1996, \apj, 470, 222
\bibitem[Smith et al. 1998]{Smith88}
	Smith, H. E., Lonsdale, C. J., Lonsdale, C. J., \& Diamond, P. J. 1998,
	\apj, 493, L17
\bibitem[Soifer et al. 1987]{Soifer87}
	Soifer, B. T., Sanders, D. B., Madore, B. F., Neugebauer, G., Danielson, G. E., Elias, J. H., Lonsdale, C. J., and Rice, W. L. 1987, \apj, 320, 238
\bibitem[Solomon et al. 1992]{Solomon92}
	Solomon, P. M., Downes, D., and Radford, S. J. E. 1992, \apj, 387, L55
\bibitem[Solomon et al. 1997]{Solomon97}
	Solomon, P. M., Downes, D., and Radford, S. J. E., and Barrett, J. W. 1997, \apj 478, 144
\bibitem[Strum et al. 1996]{Strum96}
	Sturm, E., Lutz, D., Genzel, R., Sternberg, A., Egamin, E., Kunze, D., Rigopoulou, D., Bauer, O. H., Feuchtgruber, H., Moorwood, A. F. M., and de Graauw, Th. 1996, \aap, 315, L133
\bibitem[Taniguchi 1998]{Taniguchi98}
	Taniguchi, Y., and Shioya, Y. 1998, \apjl, 501, L167
\end{thebibliography}
\end{document}